\documentclass[12pt]{article}
\usepackage{psfig,rotating}

\begin{document}

\baselineskip=25pt
\textwidth 15.0truecm
\textheight 21.0truecm
\topmargin 0.2in
\headsep 1.2cm

\title{A Peculiar Family of Jupiter Trojans: the Eurybates\footnote{Based on observations made with the Italian Telescopio Nazionale Galileo
(TNG) operated on the island of La Palma by the Fundacion Galileo Galilei
of the INAF (Istituto Nazionale di Astrofisica) at the Spanish
Observatorio del Roque de los Muchachos of the Instituto de Astrofisica
de Canarias, programmes: TAC41(AOT14) and TAC69(AOT15).}}

\author{
F. De Luise$^{1,2}$,
E. Dotto$^{1}$,
S. Fornasier$^{3,4}$,\\
M.A. Barucci$^{3}$,
N. Pinilla-Alonso$^{5,6}$,
D. Perna$^{1,3,7}$,
and F. Marzari$^{8}$
}

\maketitle

{\small
\noindent
$^1$ INAF-Osservatorio Astronomico di Roma, Via Frascati 33, 00040 Monteporzio Catone (Roma), Italy; \\
$^2$ INAF-Osservatorio Astronomico di Collurania-Teramo, Via Mentore Maggini, s.n.c., 64100 Teramo (TE), Italy; \\
$^3$ LESIA, Paris Observatory, 5 Place Jules Janssen, 92195 Meudon Cedex, France;\\
$^4$ University of Paris VII ``Denis Diderot", 4 rue Elsa Morante, 
75013 Paris, France; \\
$^5$ Fundaci\'on Galileo Galilei \& Telescopio Nazionale Galileo, PO Box 565, 38700 S/C de La Palma, Tenerife, Spain; \\
$^6$ Oak Ridge Associated Universities (ORAU) - NASA Ames Research Center, MS 245-3,
Moffett Field, CA 94035-1000;\\
$^7$ University of Rome ``Tor Vergata", Via della Ricerca Scientifica 1, 00133 Roma, Italy; \\
$^8$ Dipartimento di Fisica - University of Padova, Via Marzolo 8, 35131 Padova, Italy. \\
}

\noindent
Manuscript pages: 23 \\
Figures: 2; Tables: 2\\
\noindent
\vspace{1cm}

{\bf Running head:} Jupiter Trojans: near-infrared investigation of Eurybates Family 

\vspace{2cm}

{\it Send correspondence to:}\\
De Luise Fiore\\
INAF-Osservatorio Astronomico di Collurania-Teramo\\
Via Mentore Maggini, s.n.c.\\
I-64100 Teramo\\
Italy\\
e-mail: deluise@oa-teramo.inaf.it\\
phone: +39-0861-439712\\
fax: +39-0861-439740;\\
\\

\newpage
\vspace{2.5cm}

\begin{abstract}
\noindent

The Eurybates family is a compact core inside the Menelaus clan, located in the L$_4$ swarm of Jupiter Trojans.
Fornasier et al. (2007) found that this family exhibits a peculiar abundance of spectrally flat objects, similar to Chiron-like Centaurs and C-type main belt asteroids.
On the basis of the visible spectra available in literature,
Eurybates family's members seemed to be good candidates  for having
on their surfaces water/water ice or aqueous altered materials. \\
To improve our knowledge of the surface composition of this peculiar family, we carried out an observational campaign at the Telescopio Nazionale Galileo (TNG), obtaining near-infrared spectra of 7 members.
Our data show a 
surprisingly absence of any spectral feature referable to the 
presence of water, ices or aqueous altered materials on the surface of the 
observed objects.
Models of the surface composition are attempted, evidencing 
that amorphous carbon seems to dominate the
surface composition of the observed bodies and some amount of silicates
(olivine) could be present.

\end{abstract}


\newpage

\section{Introduction}

Jupiter Trojans (JTs) are small bodies of the Solar System located in the Jupiter's Lagrangian points L$_4$ and L$_5$.
Their origin is not yet well understood and it is still matter of debate.
Several mechanisms were proposed to model their origin (Marzari \& Scholl, 1998a,b, 2000, 2007; Marzari et al., 2002; Morbidelli et al., 2005), and it is widely accepted that
they formed in the outer Solar System, in regions rich in frozen volatiles.
The JT population is supposed to have undergone a significant collisional
evolution, and to be at least as collisionally evolved as main belt asteroids.
The discovery of dynamical families in both L$_4$ and L$_5$ clouds
supports this hypothesis
(e.g. Milani, 1993; Milani \& Kne$\breve{z}$evi\'c, 1994; Beaug\'e \& Roig, 2001; Dell'Oro et al., 1998).

Physical properties of JTs are poorly known. The presently available  spectroscopic data set is largely unsatisfactory, covering only about 10\% of the entire JT population.
To improve our knowledge of
the nature of these bodies, in the last years
several surveys have been carried out, both in visible and infrared
wavelengths (Dotto et al., 2008, and reference therein).
Although JTs formed at large heliocentric distances, the data so far 
acquired have shown
a lack of any evidence of ices on their surfaces.
 Emery \& Brown (2003, 2004) analysed the content of
water ice and
hydrated materials on the surface of 17 JTs, obtaining
upper limits of a few\% for water ice and of 30\% for
hydrated materials.
More recently, Yang \& Jewitt (2007) suggested
that water ice can occupy no more than 10\% of the total surface of
(4709) Ennomos.
JTs belonging to dynamical families do not exibit 
any spectral feature related to the presence of ices on their surface
(Dotto et al., 2006).

The data so far available in the literature put in evidence a 
great homogeneity in the whole population: 
all of the known JTs are low albedo bodies belonging to the 
primitive $C$, $P$ or $D$ classes.
The same uniformity  is also found in JTs belonging to dynamical
families (Fornasier et al., 2004, 2007; Dotto et al., 2006).
However, some differences between the L$_4$ and L$_5$ swarms are
evident: as discussed by Fornasier et al. (2007), 
the majority of L$_5$ JTs are D-types, 
while an higher presence of C and P types is observed 
among the L$_4$ objects.

A peculiar case is given by the Eurybates dynamical family, in the L$_4$ 
swarm.
This family is a strong cluster inside the Menelaus clan 
(Roig et al., 2008), 
which survives also at a very low relative velocity cut-off, as defined by Beaug\'e and Roig (2001).
The family population, up to date, is composed by 28 members at a cut-off of 70 m/s, 22 of them surviving also at 40 m/s.
On the basis of the dynamical properties, it is still not possible to
understand if the Eurybates members constitute a distinct family that
lies in the same space of proper elements of Menelaus or, as suggested by Roig et al. (2008), they formed by a secondary break-up of a former
Menelaus member.

 Although the Eurybates family is clustered, in the space of proper 
elements, in a small portion of the region occupied by the Menelaus clan, 
its members show spectral properties quite different from those of Menelaus:
as shown by Roig et al. (2008) the Menelaus clan evidences a larger 
diversity of taxonomic classes including 
C, P, and D-type objects in agreement with the 
whole JT population, while the 
Eurybates members are characterized by almost flat visible 
spectra (see e.g. Fig. 12 in Roig et al., 2008), with spectral slopes 
strongly clustered around 2 \%/10$^3$ \AA, and  spectral behaviors
similar to those of C--type main belt asteroids and/or Chiron-like Centaurs 
(Fornasier et al., 2007).

The Eurybates family assumes a great importance in the study of JTs because such a peculiar clustering of spectrally flat objects strongly affects the color-size-orbital parameter
distributions of the whole JT population investigated up to now.
Fornasier et al. (2007) noted how this family fills the distribution of 
spectrally neutral JTs at low inclination and appears to be the major  
responsible of a color-inclination trend (bluer bodies concentrated at lower 
inclination) of the whole JT population.
In the same paper, the Eurybates family appears also to be the major cause of
the abundance of C- and P-types among the L$_4$ objects,
which would imply a more
heterogeneous composition of this swarm than the L$_5$ one.
Moreover, the Eurybates family
strongly contributes to the population of L$_4$ small JTs (with a D$<$40~km)
having low spectral slopes.

The observations made by Fornasier et al. (2007) showed
the presence of a drop-off of reflectance shortward of 0.52~$\mu$m
in the visible spectra of four Eurybates members 
(18060, 24380, 24420 and 39285).
This behavior is detected on the spectra of many main belt C-type asteroids 
(Vilas, 1994; 1995), and it is often associated 
to other spectral features due to aqueous alteration products.
Since no other absorption features were found on the visible spectra 
of Eurybates members, we still do not have a final proof that aqueous 
alteration processes occurred on the surface of these bodies.
Nevertheless, the presence of the ultraviolet drop-off could suggest that
subsurface water
or water ice could have been present on Eurybates members at a certain moment 
of their life,
in order to cause aqueous alteration on their surfaces.
They are therefore good candidates to preserve still detectable spectral 
signatures of water/water ice or aqueous altered materials.

%
  \section{Observations and Data Analysis}
%

To constrain the surface composition of Eurybates family's members, 
we performed an observational
spectroscopic campaign in the near infrared (NIR) wavelength range.

The observations were carried out at the 3.6~m Telescopio Nazionale Galileo (TNG) at Roque de Los Muchachos in La Palma (Canary Islands, Spain) in 2006 and 2007 (AOT14-TAC41 and AOT15-TAC69, respectively).
The targets were selected using the list defined by Beaug\'e and Roig (2001) and the P.E.Tr.A. project\footnote{http://www.daf.on.br/froig/petra/}. 
We observed 7 objects already investigated in visible range by Fornasier et al. (2007). 
With the exception of 163135, all of our targets survive at a velocity cut-off of 40 m/s.
The observational circumstances are summarized in Tab.~\ref{tab:eury_circ}.

We used the Near Infrared Camera Spectrometer (NICS), a multimode instrument based on a HgCdTe Hawaii 1024x1024 array, with a field of view of 4.2 x 4.2 arcmin, coupled with the AMICI prism.
Our observations were carried out in low resolution spectroscopic mode, covering the 0.9--2.4 $\mu$m spectral range, using a 5 arcsec wide slit, oriented in the object moving direction.
The total exposure time was divided into several sub-spectra of 120~s each, to reduce the noise contribution typical of sky at NIR wavelengths.
The observations were done by nodding the object along the slit by 30 arcsec between two positions A and B.
Flat-fields were also acquired at the beginning of each night.

\begin{figure}[b]
\centerline{\psfig{file=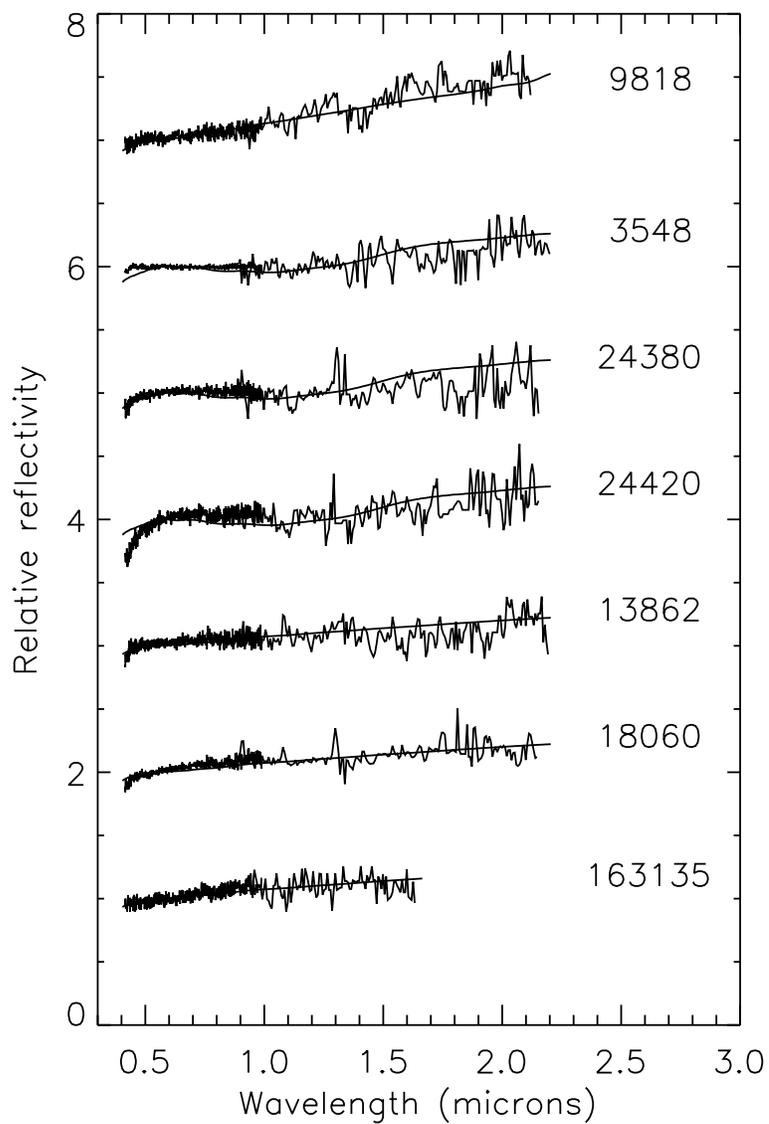,width=12truecm}}
\caption{Near-infrared spectra of Jupiter Trojans belonging to Eurybates family. All the spectra are normalized at 1.25~$\mu$m and  shifted by 1.0 in reflectance for clarity. Their mean S/N ratio value, measured at about 1.25 $\mu$m, is around 15.}
\label{fig:nir}
\end{figure}

Data were reduced using the standard procedure (e.g. Dotto et al., 2006) 
with MIDAS and the IDL software packages.
The two averaged A and B images were subtracted from each other.
The $A-B$ and $B-A$ images were flat-fielded, corrected for spatial and spectral distortion and finally combined with a 30-arcsec offset.
The spectra were hence extracted from the resulting combined images.
Wavelength calibration was obtained using a look-up table, available on the TNG website, which is based on the theoretical dispersion predicted by ray-tracing and adjusted to best fit the observed spectra of calibration sources.
The telluric absorption correction and the removal of the solar contribution 
were obtained by dividing the spectrum of each object by the spectrum of
the  
solar analog star closest in time and airmass to the target
(see Tab.~\ref{tab:eury_circ}).
The resulting spectra were smoothed with a median filtering technique, 
to reach
a spectral resolution of about 20.
The edges of each spectral region were cut to avoid low S/N regions at wavelength lower than about 0.90~$\mu$m and greater than about 2.2~$\mu$m.
Only for 163135 we cut the spectrum at 1.65~$\mu$m,
as, due to sky variability,
it was not possible to properly remove the sky contribution.
The obtained NIR spectra are shown in Fig.~\ref{fig:nir}. 

\begin{figure}[h]
\centerline{\psfig{file=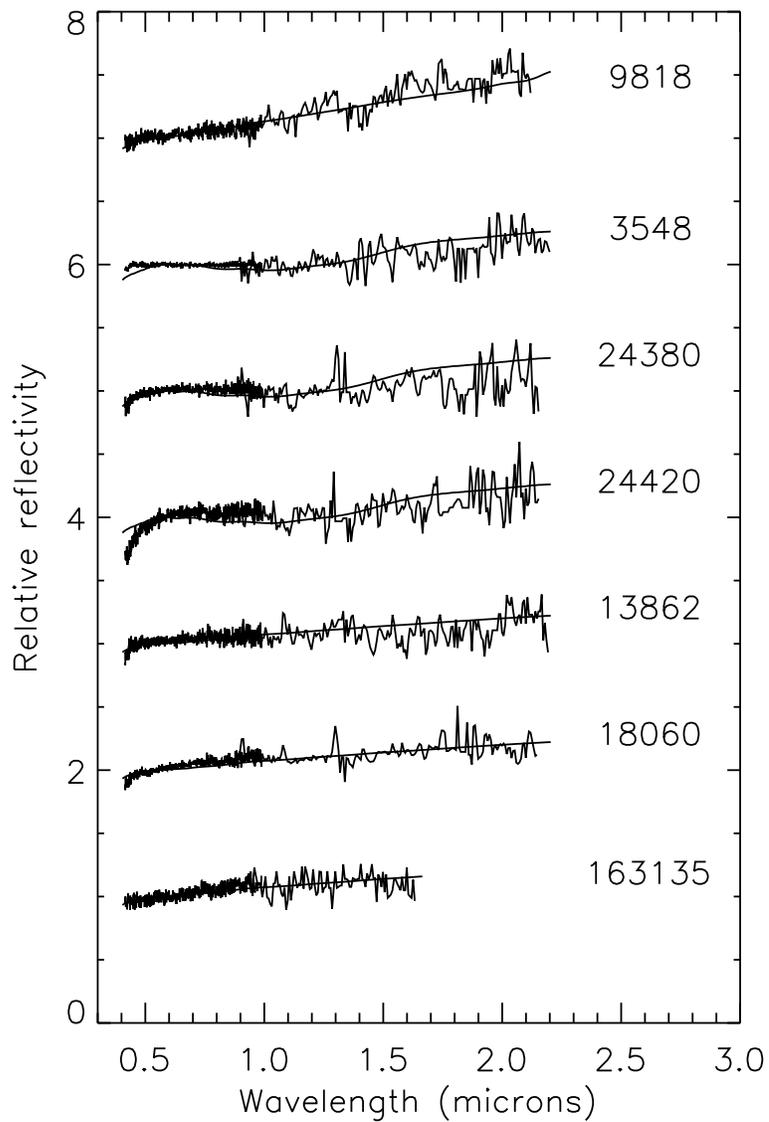,width=12truecm}}
\caption{Near-infrared spectra of Eurybates family members obtained by our observations, together with the visible part
already published by Fornasier et al. (2007). All the spectra are normalized at 0.55~$\mu$m and shifted by 1 in reflectance for clarity. The superimposed continuous lines represent the synthetic spectra obtained modeling the surface composition.}
\label{fig:eury_model}
\end{figure}

Our NIR spectra were finally combined with the visible spectra published by Fornasier et al. (2007), overlapping the common region between 0.9 and 0.95~$\mu$m. The resulting V+NIR spectra, normalized at 0.55~$\mu$m, are shown in Fig.~\ref{fig:eury_model}.
We computed the spectral slopes of all the observed objects between 1.0 and 1.6~$\mu$m (see  Tab.~\ref{tab:eury_mod}).  
The obtained values span a small range of values, from 0.11 to 3.60 \%/10$^3$\AA, with a mean value of $1.43 \pm 0.41$ \%/10$^3$\AA.

The obtained spectral behaviors allow us to confirm the taxonomic 
classification published by Fornasier et al. (2007) 
 (see Tab.~\ref{tab:eury_mod}). 
Our
investigation in NIR wavelengths has surprisingly shown featureless spectra:
we did not detect any spectral feature between 0.9 and 2.2~$\mu$m referable to the presence, on the surface of the observed bodies, of  water, ices or 
hydrated minerals.

To model the surface composition of the observed Eurybates members
we used the radiative transfer model, based on the Hapke theory, already applied to JTs by Dotto et al. (2006).
We took into consideration the following materials:
amorphous carbon (by Zubko et al., 1996), 
organic solids (e.g. kerogens by Clark et al., 1993 and Khare et al., 1991; Triton tholins by McDonald et al., 1994;
titan tholins (by Khare et al., 1984) 
all the minerals present in the RELAB database\footnote{http://www.planetary.brown.edu/relabdocs/relab\_disclaimer.htm},
bitumen (by Moroz et al., 1998), and ices (H$_2$O, CH$_4$, CH$_3$OH, NH$_3$, and ice tholin by McDonald et al., 1996 and Khare et al., 1993).
To model the surface composition of each observed object, we considered 
several geographical mixtures of all these compounds.
For each mixture, the modeling procedure produced a synthetic spectrum, 
to be 
compared with the observed one, 
and calculated the geometric albedo value at 0.55~$\mu$m. A $\chi^2$-test 
was applied to compare the different models tentatively considered for each 
target, and to select the model which better reproduces the observed spectrum. 
In this analysis we did not take into account the critical regions of the
spectrum, around 1.4 and 1.9 $\mu$m, where telluric bands occur.
We considered as best model the geographycal
mixture best fitting the asteroid spectrum, and having an albedo value
compatible 
with the typical value
of C- or P-type dark asteroids and the mean value for JTs 
(0.041$\pm$0.002, as computed by Fern\'andez et al., 2003).
In Fig.~\ref{fig:eury_model} the synthetic spectra of final models (continuous lines) are superimposed on the observed spectra.
Table~\ref{tab:eury_mod} reports, for each object, the model of surface composition, as well as the computed albedo.

The obtained spectra suggest the predominance of amorphous carbon on the surface of the observed members of the Eurybates family. 
In particular, the spectra of 13862, 18060 and 163135 are 
similar to the one of pure amorphous carbon.
The spectral behaviors of (3548) Eurybates, 24380 and 24420 suggest the presence on their surface of a few amount of olivine.
The slightly spectral reddening of (9818) Eurymachos has been modelled using a small percentage of a reddening agent (e.g. Triton tholin).

%
  \section{Discussion and conclusions}
%

All the spectra of Eurybates members presented in this work appear 
flat and featureless and 
confirm the taxonomic classification published by Fornasier et al. (2007).
Our 
surface modeling has shown that the amorphous carbon seems to dominate the 
surface composition of the observed bodies and some amount of silicates 
(olivine) could be 
present.
The proposed models are not unique, since they depend on many
parameters (e.g. physical properties of the surface, optical constants and
particle size), 
but a complete lack of diagnostic features typical of
water, ices and hydrated minerals is evident in our spectra. 
This result does not allow us to definitively exclude that
some percentage of water ice is still present
on the observed bodies, hidden by dark materials.
Brunetto \& Roush (2008) showed 
that few tens of microns of an 
organich-rich layer (e.g. irradiated methan ice), produced by space weathering, 
are enough to mask spectroscopically, in the 
near-infrared wavelength range, the presence of water and other 
volatiles below the surface.

The spectral evidences presented in this paper leave open several possibilities about the origin of the Eurybates family.

A first scenario implies the formation of the Eurybates family by
the disruption of an exogenous body, i.e. coming from other Solar System
regions, probably captured by Jupiter gravitational field and trapped in L$_4$ Lagrangian point.
In this case, the origin of the parent body is a crucial point that
must yet be assessed, as well
as the nature and the efficiency of the capture mechanism.
Of course, it is plausible that this captured parent body is not the only one in the Trojan clouds, but
the population of these objects must be still assessed.

Other scenarios take into account the action of space weathering processes,
still efficient at 5.2 AU from the Sun where JTs are
presently orbiting (see e.g. Strazzulla et al. 2005; Melita et al. 2009).
We know, from laboratory experiments, that the effect of ageing mechanisms
strongly depends on the composition and nature of the surfaces
exposed to space weathering.
Since we still do not know  the origin and primordial composition of JTs, 
and 
therefore we do not know how space weathering processes acted on JT surfaces, 
several scenarios have to be considered:\\
a) if JTs had icy surfaces, the space weathering processes would have 
produced
an irradiation mantle spectrally red and with low albedo (Moore et al.,
1983; Thompson et al., 1987; Strazzulla, 1998; Hudson \& Moore, 1999; Brunetto et al., 2006; Brunetto \& Roush, 2008); \\
b) a similar result would have been produced on silicatic composition, 
where space
weathering produces a gradual spectral reddening, as already observed 
in several dynamical families in
the asteroid main belt (e.g. the Eunomia family by Lazzaro et al., 1999;
the Flora clan by Florczak et al., 1998; and
the Eos family by Doressoundiram et al., 1998) and shown by laboratory
experiments (e.g. Strazzulla et al. 2005; Lazzarin et al. 2006);\\
c) an opposite result would have been produced on a surface covered of
natural complex hydrocarbons, where ion irradiation would have produced 
gradually neutralized spectra (Moroz et al., 2004).

\noindent The first two cases
bring to the possibility that the Eurybates is a young family, 
produced either by a the fragmentation of an object coming from outside
the present Trojan population and trapped around L$_4$, or by a
secondary collision involving one of Menelaus' family members.

\noindent Under the scenario \textit{(c)}, Eurybates should be an old family, with an initial hydrocarbon composition,
on which space weathering flattened the members' spectra,
wiping out the primordial differences.

\noindent Whether JTs experienced a phase of cometary activity,
water ice on their surface could have been devolatized, or
they could have formed a thin dust mantle
as shown by Rickman et al. (1990)
and Tancredi et al. (2006).
This mechanism is quite probable for large JTs not belonging to dynamical
families, and it is still plausible
for members of dynamical families, since we cannot exclude that they
suffered some episodic cometary activity
during their life after the fragmentation of the parent body.
In a recent work Melita et al. (2009) estimated the timescales of
the sublimation of amorphous water ice, the collisional resurfacing, and
the flattening of the spectral slopes by solar irradiation in the region
of JTs.
According to these authors, a dust layer, probably with
a red spectroscopic slope, does exist on the surface of JTs
as a result of the rapid sublimation of water ice after resurfacing impact
events. If this dust layer remains unaltered for more than $10^3$ years,
its spectroscopic slope is flattened by the action of solar protons.
According to the estimated timescales, impacts are so frequent that
the irradiation mantle is usually disrupted for JTs, but in
the case of the Eurybates family the flat spectra would suggest that we are seeing aged surfaces.

The knowledge of
the age of the Eurybates family is hence fundamental for understanding the
parent body origin and nature. More observations are also absolutely needed to
see whether this peculiar family is unique or if more Eurybates-like 
families
are present in both L$_4$ and L$_5$ swarms. \\
At the same time more observations of Menelaus objects are
needed to compare Eurybates and Menelaus members with the whole
population of JTs, to have more hints on the relation between 
Eurybates family and Menelaus
clan, and to cast some light on their (common or not)
origin.

\bigskip

%
     {\bf Acknowledgments} \\
%
FDL want to thank the TNG support astronomers and telescope operators, in particular W. Boschin and G. Tessicini, for their support and for the kind help to optimize the observations.
The authors thank G.A. Baratta for the helpful discussions.
The authors thank as well both the anonymous reviewers for providing comments and suggestions helpful to improve the manuscript.

\bigskip


\newpage

{\bf References}

- Beaug\'e, C. \& Roig, F., 2001. A Semianalytical Model for the Motion of the Trojan Asteroids: Proper Elements and Families, \textit{Icarus} 153, 391-415

- Brunetto \& Roush, 2008. Impact of irradiated methane ice crusts on compositional interpretations of TNOs
(Research Note). \textit{Astron. Astrophys} 481, 879-882

- Brunetto, R., Barucci, M. A., Dotto, E., Strazzulla, G., 2006. Ion Irradiation of Frozen Methanol, Methane, and Benzene: Linking to the Colors of Centaurs and Trans-Neptunian Objects, \textit{Astroph. J.} 644, 646-650

- Clark, R.N., Swayze, G.A., Gallagher, A.J., King, T.V.V., Calvin, W.M., 1993. U.S. Geological Survey Open File Report 93-592, http://speclab.cr.usgs.gov.

- Dell'Oro, A., Marzari, P., Paolicchi, F., Dotto, E., Vanzani, V., 1998. Trojan collision probability: A statistical approach. \textit{Astron. Astrophys} 339, 272-277.

- Dotto, E., Fornasier, S., Barucci, M.A., Licandro, J., Boehnhardt, H., Hainaut, O., Marzari, F., de Bergh, C., De Luise, F., 2006. The surface composition of Jupiter Trojans: Visible and Near-Infrared Survey of dynamical families. \textit{Icarus} 183, 420-434.

- Dotto, E., Emery, J.P., Barucci, M.A., Morbidelli, A., Cruikshank, D.P., 2008. De Troianis: The Trojans in the Planetary System. In: \textit{The Solar System Beyond Neptune} (M. A. Barucci, H. Boehnhardt, D. P. Cruikshank, and A. Morbidelli Eds.), Univ. of Arizona Press, Tucson, 592, 383-395

- Doressoundiram, A., Barucci, M.A., Fulchignoni, M., Florczak, M., 1998. EOS Family: A Spectroscopic Study. \textit{Icarus} 131, 15-31

- Emery, J.P. \& Brown, R.H., 2003. Constraints on the surface
composition of Trojan asteroids from near-infrared (0.8-4.0
$mu$m) spectroscopy. \textit{Icarus} 164, 104-121.

- Emery, J.P. \& Brown, R.H., 2004. The surface composition of
Trojan asteroids: Constraints set by scattering theory. \textit{Icarus} 170, 131-152.

- Fern\'andez, Y.R., Sheppard, S.S., Jewitt, D.C., 2003. The Albedo Distribution of Jovian Trojan Asteroids. \textit{Astron. J.}  126, 1563-1574

- Florczak, M., Barucci, M.A., Doressoundiram, A., Lazzaro, D., Angeli, C.A., Dotto, E., 1998. A Visible Spectroscopic Survey of the Flora Clan \textit{Icarus} 133, 233-246

- Fornasier, S., Dotto, E., Marzari, F., Barucci, M.A., Boehnhardt, H., Hainaut, O., de Bergh, C., 2004. Visible and near-infrared spectroscopic survey of Jupiter Trojan asteroids: investigation of dynamical families, 172, 221-232

- Fornasier, S., Dotto, E., Hainaut, O., Marzari, F., Boehnhardt, H., De Luise, F., Barucci, M. A., 2007. Visible spectroscopic and photometric survey of Jupiter Trojans: Final results on dynamical families. \textit{Icarus} 190, 622-642

- Khare, B.N., Sagan, C., Arakawa, E.T., F. Suits, Callcott, T.A., Williams, M.W., 1984. Optical constants of organic tholins produced in a simulated titanian atmosphere - From soft X-ray to microwave frequencies.  \textit{Icarus} 60, 127-137.

- Khare, B.N., Thompson, W.R., Sagan, C., Arakawa, E.T., Meisse, C., Gilmour, I., 1991. Optical constants of kerogen from 0.15 to 40 micron: Comparison with meteoritic organics. In: \textit{Origin and Evolution of Interplanetary Dust}, IAU Colloq. 126 (Levasseur-Regours, A.C., Hasegawa, H. Eds.),  Kluwer Academic, Dordrecht. \textit{ASSL} 173, 99.

- Khare, B.N., Thompson,  W.R., Cheng, L., Chyba, C., Sagan,  C.,
Arakawa,  E.T., Meisse,  C., Tuminello, P.S., 1993. Production and optical constraints of ice tholin from charged particle irradiation of (1:6) C2H6/H2O at 77 K. \textit{Icarus} 103, 290-300.

- Hudson, R.L. \& Moore, M.H., 1999. Laboratory studies of the formation of methanol and other organic molecules by water + carbon monoxide radiolysis: Relevance to comets, icy satellites, and interstellar ices. \textit{Icarus} 140, 451-461.

- Lazzarin, M., Marchi, S., Moroz, L. V., Brunetto, R., Magrin, S., Paolicchi, P., Strazzulla, G., 2006. Space Weathering in the Main Asteroid Belt: The Big Picture. \textit{ApJ} 647, 179-182

- Lazzaro, D., Moth\'e-Diniz, T., Carvano, J.M., Angeli, C.A., Betzler, A.S., Florczak, M., Cellino, A., Di Martino, M., Doressoundiram, A., Barucci, M.A., Dotto, E., Bendjoya, P., 1999. The Eunomia Family: A Visible Spectroscopic Survey. \textit{Icarus} 142,  445-453

- Marzari, F. \& Scholl, H., 1998a. Capture of Trojans by a growing proto-Jupiter. \textit{Icarus} 131, 41-51.

- Marzari, F. \& Scholl, H., 1998b. The growth of Jupiter and Saturn and the capture of Trojans. \textit{Astron. Astrophys.} 339, 278-285.

- Marzari, F. \& Scholl, H., 2000 The Role of Secular Resonances in the History of Trojans. \textit{Icarus} 146, 232-239.

- Marzari, F. \& Scholl, H., 2007. Dynamics of Jupiter Trojans during the 2:1 mean motion resonance crossing of Jupiter and Saturn.  \textit{Mon. Not. R. Astron. Soc.} 380, 479-488

- Marzari, F., Scholl, H., Murray, C., Lagerkvist, C., 2002. Origin and evolution of Trojan asteroids. In: \textit{Asteroids III} (Bottke Jr., W.F., Cellino, A., Paolicchi, P., Binzel, R.P. Eds.). Univ. of Arizona Press, Tucson, 725-738.

- McDonald, G.D., Thompson,W.R., Heinrich, M., Khare, B.N., Sagan, C., 1994. Chemical investigation of Titan and Triton tholins. \textit{Icarus} 108, 137-145.

- McDonald, G.D., Whited, L.J., Deruiter, C., Khare, B.N.,
Patnaik,  A., Sagan, C., 1996. Production and chemical analysis of cometary ice tholins. \textit{Icarus} 122, 107-117.

- Melita, M. D., Strazzulla, G., Bar-Nun, A., 2009, Collisions, Cosmic Radiation and the Colors of the Trojan Asteroids. \textit{Icarus} 203, 134-139.

- Milani, A., 1993. The Trojan asteroid belt: Proper elements, stability, chaos and families. \textit{Celestial Mechanics and Dynamical Astronomy} 57, 59-94.

- Milani, A., and Kne$\breve{z}$evi\'c, Z., 1994. Asteroid proper elements and the dynamical structure of the asteroid main belt.  107, 219-254.

- Moore, M.H., Donn, B., Khanna, R., A'Hearn, M.F., 1983. Studies of protonirradiated cometary-type ice mixtures. \textit{Icarus} 54, 388-405.

- Morbidelli, A., Levison, H.F., Tsiganis, K., Gomes, R., 2005.
Chaotic capture of Jupiter's Trojan asteroids in the early Solar
System. \textit{Nature} 435, 462-465.

- Moroz, L.V., Arnold, G., Korochantsev, A.V., Wasch, R., 1998. Natural solid bitumens as possible analogs for cometary and asteroid organics. \textit{Icarus} 134, 253-268

- Moroz, L., Baratta, G., Strazzulla, G., Starukhina, L., Dotto, E., Barucci, M.A., Arnold, G., Distefano, E., 2004. Optical alteration of complex organics induced by ion irradiation. 1. Laboratory experiments suggest unusual space weathering trend. \textit{Icarus} 170, 214-228.

- Rickman, H., Fern\'andez, J. A., Gustafson, B. A. S., 1990. Formation of stable dust mantles on short-period comet nuclei
\textit{Astron. Astrophys.} 237, 524-535.

- Roig, F., Ribeiro, A.O., Gil-Hutton, R., 2008. Taxonomy of asteroid families among the Jupiter Trojans: comparison between spectroscopic data and the Sloan Digital Sky Survey colors.  \textit{Astron. Astrophys.} 483, 911-931.

- Strazzulla, G., 1998. Chemistry of ice induced by bombardment with energetic charged particles. In: \textit{Solar System Ices International Symposium} (Schmitt, B., de Bergh, C., Festou, M. Eds.), Kluwer Academic Dordrecht, \textit{ASSL} 227, 281.  

- Strazzulla, G., Dotto, E., Binzel, R., Brunetto, R., Barucci, M. A., Blanco, A., Orofino, V., 2005. Spectral alteration of the Meteorite Epinal (H5) induced by heavy ion irradiation: a simulation of space weathering effects on near-Earth asteroids. \textit{Icarus} 174, 31-35.

- Tancredi, G., Fern\'andez, J. A., Rickman, H., Licandro, J., 2006. Nuclear magnitudes and the size distribution of Jupiter family comets.  \textit{Icarus} 182, 527-549.

- Thompson,W.R., Murray, B.G.J.P.T., Khare, B.N., Sagan, C., 1987. Coloration and darkening of methane clathrate and other ices by charged particle irradiation - Applications to the outer Solar System. \textit{J. Geophys. Res.} 92, 14933-14947.

- Vilas, F., 1994. A quick look method of detecting water of hydration in small Solar System bodies. \textit{Lunar Planet. Sci.} 25, 1439-1440.

- Vilas, F., 1995. Is the U-B color sufficient for identifying water of 
hydration on solar system bodies? \textit{Icarus} 115, 217-218.

- Yang, B. \& Jewitt, D., 2007. Spectroscopic Search for Water Ice on Jovian Trojan Asteroids. \textit{Astron. J.} 134, 223-228.

- Zubko, V.G., Mennella, V., Colangeli, L., Bussoletti, E., 1996. Optical constants of cosmic carbon analogue grains. I. Simulation of clustering by a modified continuous distribution of ellipsoids. \textit{Mon. Not. R. Astroph. Soc.} 282, 1321-1329.

\newpage

%
%

\newpage

{\bf TABLES and FIGURES }

\begin{table*}[ht]
       \begin{center}
       \caption{Observational circumstances. For each object, we report observing date, exposure time, airmass and solar analog used (with correspondent airmass).}
        \label{tab:eury_circ}
\begin{tabular}{ |l|c|c|c|c|}
\hline
\hline

Object  &  Date &  T$_{exp}$ & Airmass & Solar Analog \\
        &       & ($s \times$ n$_{acq}$)	  && (airmass)      \\
\hline

3548 Eurybates  & 18 Aug 2006  &  120 x 8  & 1.8     & Hip102491 (1.9) \\
13862           & 15 Jul 2007  &  120 x 16 & 1.6     & HD210078  (1.4) \\
18060           & 16 Jul 2007  &  120 x 16 & 1.3     & HD210078  (1.3) \\
9818 Eurymachos	& 17 Jul 2007  &  120 x 24 & 1.7     & HD210078  (1.4) \\
24380           & 17 Jul 2007  &  120 x 16 & 1.4     & HD210078  (1.3) \\
24420           & 04 Aug 2007  &  120 x 16 & 1.5     & HD210078  (1.3) \\
163135          & 04 Aug 2007  &  120 x 32 & 1.5-1.8 & HD210078 (1.3)  \\

\hline

\end{tabular}
\end{center}
\end{table*}

\begin{table*}[ht]
       \begin{center}
       \caption{For each object, the spectral slope $S_{NIR}$ (computed between 1.0 and 1.6~$\mu$m), taxonomic classification given by Fornasier et al. (2007), 
the model of the surface composition and computed albedo value at 0.55 $\mu$m are reported. The used acronyms are:  AC = amorphous carbon, Tr. th. = Triton tholin, Ol = olivine.}
        \label{tab:eury_mod}
\begin{tabular}{ |l|c|c|c|c|c|}
\hline
\hline

Object          &       $S_{NIR}$  & Tax.  & Model & Albedo \\
                &  (\%/10$^3$\AA)  & class &       &        \\
\hline

3548 Eurybates  &  $1.82 \pm 0.41 $ &  C   & 99\% AC -- 1\% Ol     & 0.03  \\
13862           &  $0.11 \pm 0.31 $ &  C   & 100\% AC  & 0.03\\

18060           &  $1.11 \pm 0.27 $ &  P   & 100\% AC  	   & 0.03  \\

9818 Eurymachos	&  $3.60 \pm 0.38 $ &  P   & 98\% AC -- 2\% Tr. th.  & 0.03 \\

24380           &  $0.98 \pm 0.55 $ &  C   & 99\% AC -- 1\% Ol     & 0.03  \\
24420           &  $2.12 \pm 0.50 $ &  C   & 99\% AC -- 1\% Ol     & 0.03  \\
163135          &  $0.25 \pm 0.55 $ &  P   & 100\% AC	           & 0.03  \\

\hline

\end{tabular}
\end{center}
\end{table*}

\newpage

%
%

\end{document}